\documentclass[11pt,prl,preprint]{revtex4}   
\usepackage[dvips]{epsfig}   
\usepackage[dvips]{graphicx}   
\usepackage{amsmath,amssymb,amsfonts}   
\usepackage{subfigure}   
\usepackage{revsymb}   

\newcommand{\LN}{ ${\rm LiNbO_3}\;$}

\begin{document}
   
\title{Harmonic generation in a two dimensional nonlinear quasicrystal}   
\author{R.~T.~Bratfalean}
\affiliation{Optoelectronics Research Centre, University of Southampton,   
Southampton, SO17 1BJ UK}   
\author{A.C.~Peacock}
\affiliation{Optoelectronics Research Centre, University of Southampton,   
Southampton, SO17 1BJ UK}   
\author{Ruth Lewen}
\affiliation{Electronics and Computer Science, University of Southampton,   
Southampton, SO17 1BJ UK}   
\author{N.~G.~R.~Broderick}
\email{ngb@orc.soton.ac.uk}
\affiliation{Optoelectronics Research Centre, University of Southampton,   
Southampton, SO17 1BJ UK}   
\author{K.~Gallo}
\affiliation{Optoelectronics Research Centre, University of Southampton,   
Southampton, SO17 1BJ UK}   
   
\begin{abstract}   
Second harmonic generation in a two dimensional nonlinear quasi-crystal
is demonstrated for the first time. Temperature and wavelength
tuning of the crystal reveal the uniformity of the pattern while angle
tuning reveals the dense nature of the crystal's Fourier spectrum.
These results compare well with theoretical predictions showing the 
excellent uniformity of the crystal and suggest that more complicated
``nonlinear holograms'' should be possible.
\end{abstract}   

\maketitle

Periodic poling of ferroelectric materials such as lithium niobate
is a well established technique for improving the efficiency of
nonlinear optical processes\cite{qpm1}. When combined with channel
waveguide technology, periodically poled lithium
niobate (PPLN) is perhaps the premier system for nonlinear optics at
telecommunications wavelengths. Recently work has begun in
exploring both theoretically\cite{berger,norton1} and
experimentally\cite{hexln,winter} the properties of two
dimensionally poled nonlinear photonic crystals.  We present here
the first example of  a two dimensionally poled nonlinear
quasi-crystal and discuss its properties.

The importance of a PPLN crystal can best be understood by
considering second harmonic generation (SHG). In this process two
photons of frequency $\omega_1$ are converted into one photon of
frequency $2 \omega_1$ a process that automatically conserves
energy. For this interaction to occur momentum must also be conserved
leading to: 
\begin{equation}
\label{momentum}
    {\bf k}(\omega) + {\bf k}(\omega) +{\bf G}_n = {\bf k}(2 \omega)
\end{equation}
where ${\bf k}$ is the wavevector of the light in the material and ${\bf G}_n$ 
is a reciprocal lattice vector of the crystal. For a strictly
periodic PPLN crystal $|{\bf G}_n| \propto n/d$ where $n \in {\mathbb Z}$
and $d$ is the period of the crystal. As the period of the
crystal can be freely chosen during the fabrication process, 
Eq.~\eqref{momentum} is also a design rule for PPLN
allowing the phase matching of  any {\it single} interaction.

If we consider a harder problem that of simultaneously phase matching
two nonlinear processes we can immediately see that instead of having
to satisfy one momentum conservation equation we need to satisfy two. 
With the only free parameter in a PPLN crystal being the domain period $d$ 
this cannot in general be done as the phase mismatches are rarely integer 
multiples of
some constant. Hence an alternative approach must be taken. The
simplest scheme would be to write two gratings on top of one another,
however due to the digital nature of the poling process (essentially the
nonlinear coefficient can be either $\pm 1$) and the fact that there is
a minimum domain size of $\approx 5\, \mu {\rm m}$, means that this is
not possible and more complicated patterns must be poled. Two different
patterns have been suggested in the literature. The first is to use
an aperiodic pattern usually based on a Fibonacci sequence\cite{fib} 
while the second is to use a two dimensional periodic pattern\cite{saltiel}. 
Our aim in this paper is to look at the combination of 
both approaches, i.e. a two dimensional aperiodic poled pattern.

\begin{figure}[h]
\centerline{
\psfig{file=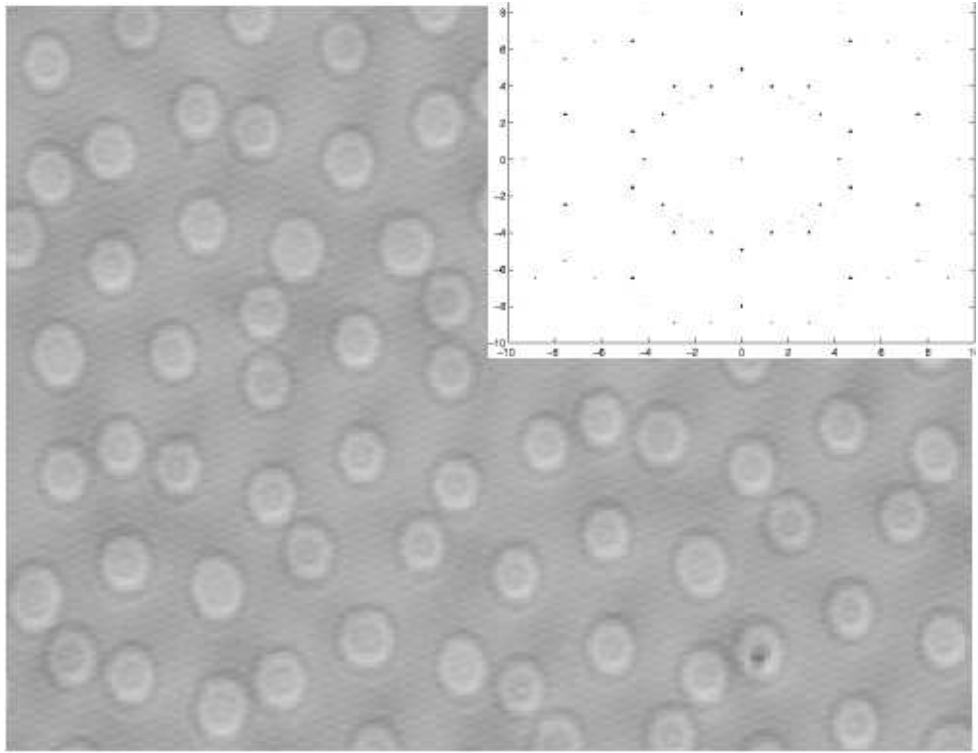,height=10cm}
}
\caption{\label{pattern} Optical picture of the poled pattern used
in the experiments. The insert shows the Fourier transform of the 
pattern. }
\end{figure}

%
More generally there is a need to produce poled patterns that are not
strictly periodic such as chirped PPLN, superstructured PPLN, transversed
pattern PPLN etc\cite{fejer1d,fejer2d,fejer2d2}. In all these cases the additional 
structure in real space provides
additional functionality (e.g. pulse compression, spatial focussing, etc.) 
and we can consider these gratings as simple cases of ``nonlinear holography" 
in which both the temporal and frequency components of an input pulse are 
altered to produce an output pulse with both a different central frequency and 
pulse shape. Doing this in an arbitrary fashion would require the ability to 
pole arbitrary shapes in a 2D crystal and as a stepping stone we look in this 
paper at the poling and optical response of a 2D Penrose pattern as shown in 
Fig.~\ref{pattern}.


A Penrose pattern is an aperiodic tiling of the plane which has long range
order but which lacks any translational or rotational symmetry\cite{penrose}. 
As shown in
Fig.~\ref{pattern}, a Penrose pattern can be thought of as a tiling of
a plane by two rhombuses - a ``thick" rhombus and a ``thin'' rhombus. 
 To go from
the Penrose lattice to our crystal structure we decided to place a poled
region at each vertex point (the alternative of placing a poled region at
the centre of each rhombus was considered and makes no significant 
difference). Upon taking the Fourier transform of the resulting structure
the result is as shown in Fig.~\ref{pattern} (insert). In the insert the 
location of the dots correspond to the location of reciprocal lattice vectors
which can be used for quasi-phase matching [via Eq.~\eqref{momentum}]
while the size of the dots 
indicate the size of the corresponding Fourier coefficients.
    
As there are an infinite number of nonequivalent Penrose tilings of the
plane\cite{penrose2}
(two tilings are nonequivalent if one cannot be transformed into
the other by a simple translation and rotation) it was necessary to 
decide which one to use. Clearly initially the pattern must be scaled
so that the reciprocal lattice vectors satisfy Eq.~\eqref{momentum}. Then
the size and the shape of the poled regions must be chosen to maximum
the relevant Fourier coefficient ( in a QPM process the strength of the
interaction is proportional to the size of the Fourier coefficient 
corresponding to the reciprocal lattice vector mediating the process).
We numerically generated several different patterns and optimised the
shape of the poled region resulting in a pattern that could be used to 
phase match second harmonic generation at
$140^{\rm o}\,{\rm C}$ with a fundamental wavelength of 1536nm.

We then wrote a 20.53mm by 1.15mm portion of this Penrose pattern onto
a lithography mask which was used to pole the \LN crystal in the standard
way.  A thin  layer of photo-resist was deposited on the -z face of the 
0.5mm thick, z-cut LiNbO3  sample and then photo-lithographically patterned 
according to the mask. The width and lengths of the Penrose regions were 
carefully aligned with the x and y crystallographic axes of \LN, respectively. 
Poling was accomplished by applying an electric field  via liquid electrodes 
on the $\pm z$ faces at room temperature. As shown in Fig.~\ref{pattern},  
the Penrose pattern fabricated onto the sample was found to be uniform 
across the  entire poled area and was faithfully reproduced throughout the 
crystal depth (z axis). Finally we polished the  $\pm x$ faces, 
allowing a propagation length of 20mm through the sample.
  
\begin{figure}[h]
\centerline{
\psfig{file=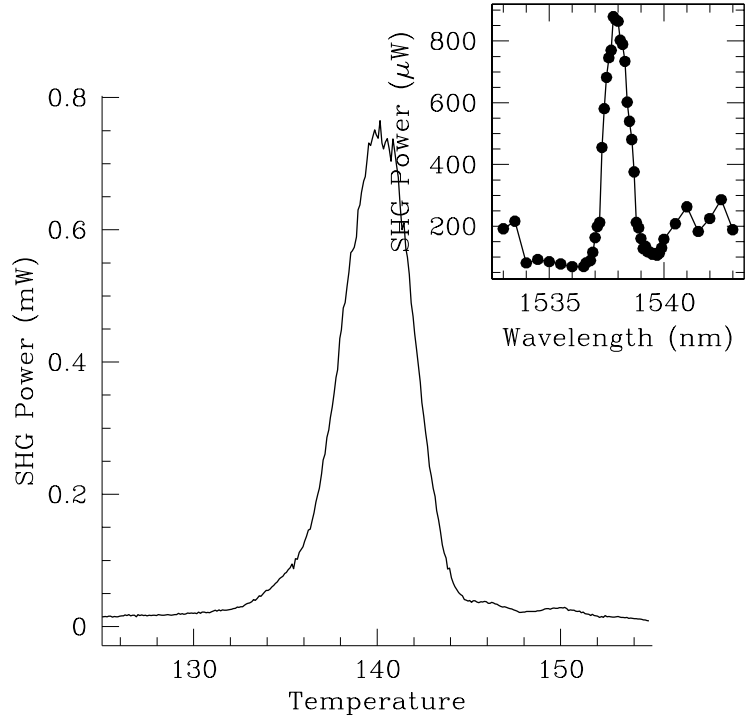,height=10cm}
}
\caption{\label{temp} Temperature Tuning curve for the Penrose Pattern.
The insert shows the wavelength tuning response.  }
\end{figure}

We first measured the temperature tuning curve for second
harmonic generation. The crystal was rotated with respect to 
the incident beam to obtain maximum SHG at $140\, ^{\rm o}{\rm C}$. 
Next the temperature tuning curve was measured over the range
$120 - 160 \, ^{\rm o}{\rm C}$ as shown in Fig.~\ref{temp}  and has a FWHM of 
$5.1\,^{\rm o}{\rm C}$. For this measurement the incident fundamental
beam was gently focussed into the crystal and consisted of 5 ns pulses at a
repetition rate of 100 kHz and an average  power of 300 mW which was 
sufficiently small to ensure we were in the low conversion regime. 
The input beam was confocallly focused to a spot
size of $162\,\mu {\rm m}$ in the centre of the crystal.  Due to 
non-collinear nature of the interaction we would expect a Gaussian 
response rather than a sinc shaped one\cite{hexln2} and that it should
be broader than a collinear response. 

Next we measured the wavelength response of the crystal over the range
1533nm to 1543nm using a tuneable source with an average power of 
150mW. The crystal was rotated to give maximum SHG at 1538nm and kept at 
a constant temperature of $140\, ^{\rm o}{\rm C}$. The results are 
as shown in Fig.~\ref{temp} (insert). We found a FWHM of 1.3nm which
corresponds well to the measured temperature tuning width. 

\begin{figure}[h]
\centerline{
\psfig{file=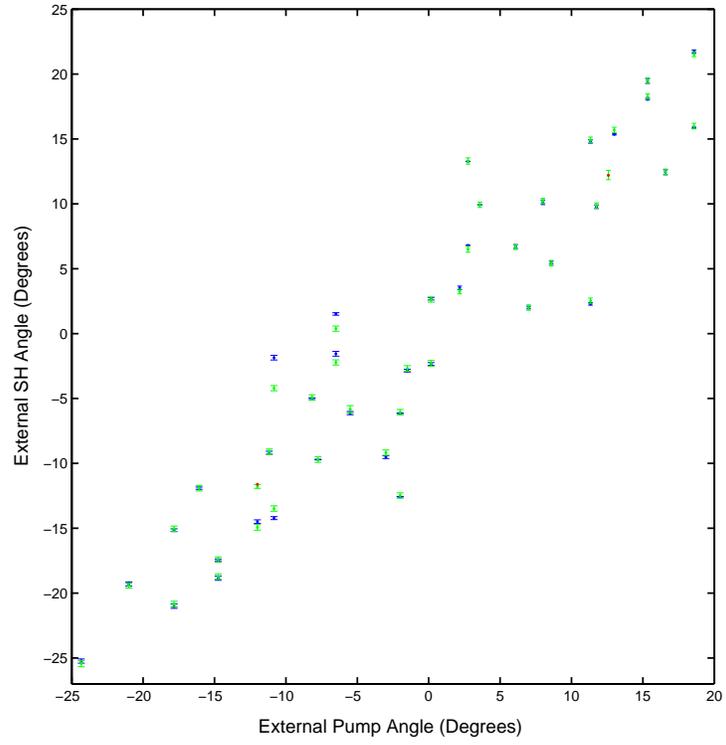,height=10cm}
}
\caption{\label{angles} Picture of the measured and theoretical input and
output angles for the Penrose pattern.}
\end{figure}

A distinguishing feature of an aperiodic crystal compared to a periodic crystal
is the dense nature of its Fourier transform. As such it should be
possible to observe SHG with our crystal for a much larger set of input angles
than for a periodic crystal (see Fig.~4 in \cite{hexln}). To do this 
we fixed the input wavelength and incident power and the crystal temperature
and rotated the crystal noting when a SH spot appeared. For each
SH beam that was produced we measured the input angle of the fundamental
along with the angle between the fundamental and second harmonic. These
results are shown in Fig.~\ref{angles} (circles). We measured 43 distinct
spots which was roughly 50\% more that what we had measured previously
for a periodic crystal (the numbers are not strictly comparable as the
input power was roughly 1000 times bigger in the earlier 
measurement\cite{hexln}!) illustrating the dense nature of the Fourier
space. In addition we have shown on Fig.~\ref{angles} the theoretically
predicted input and output angles for the Fourier peaks involved in the
interactions. 

The importance of measuring the output SHG spots is that it directly
gives us information about the position and strength of the peaks in
Fourier space which in turn gives us information about the crystal
lattice. For example Fig.~\ref{angles} is mirror symmetric to a good
degree indicating a mirror plane in Fourier space. We were not
however able to rotate the crystal sufficiently to observe fully the
5 five symmetry but the beginnings of it can be seen. We are able to 
account for almost all of the experimental spots apart from two
while similarly only one theoretical Fourier peak appears for which there
is no corresponding experimental spot. The reason for this is not currently
known although we believe that it might relate to overpoling of the 
crystal in some regions causing a change in the size of the Fourier
coefficients.

Finally we measured the efficiency of the second harmonic generation
for near normal incidence and at a temperature of $105^{\rm o}\,{\rm C}$.
The reason for this is that while angle tuning of the crystal gives us
information about the location of the peaks in Fourier space it does not
provide us with any information about their size. However the SH power is
proportional to the square of the Fourier coefficients which in turn
depends on the size of the poled regions. For the spot we chose we 
measured an efficiency of 1.64\% while the predicted theoretical 
efficiency for a collinear interaction would be 2.73\%. The two most
likely causes for this discrepancy are the noncollinear nature of our
interaction which would reduce the efficiency and the fact that we do
not precisely know the size of the poled regions over the whole of the
crystal. Thus our measurements suggest that there is some degree of
overpoling in the crystal. We note that although the efficiency is low
this is due to the choice of Fourier peak choosing a different Fourier 
peak should in principle allow conversion efficiencies $> 90\%$ for 
our available pump power.
   
In conclusion we have designed and fabricated the first two dimensional
nonlinear photonic quasicrystal for use in second harmonic generation. 
Measurements on this crystal demonstrate the greater density of 
reciprocal lattice vectors compared to a periodic crystal along with the
five fold rotational symmetry. Such crystals are likely to find use in
the simultaneous phase-matching of multiple nonlinear interactions due to their
flexibility in the location of the Fourier peaks. The fabrication of such
a non-periodic crystal is a significant step forwards towards the design
and fabrication of ''nonlinear holograms" which would allow complete
spatial and temporal shaping of input beams through the nonlinearity.

\bibliographystyle{osa}

\begin{thebibliography}{10}

\bibitem{qpm1}
J.~A. Armstrong, N. Bloembergen, J. Ducuing, and P.~S. Pershan, ``Interactions
  between Light Waves in a Nonlinear Dielectric,'' Phys. Rev. {\bf 127,} 1918
  (1962).

\bibitem{berger}
V. Berger, ``Nonlinear Photonic Crystals,'' Phys. Rev. Lett. {\bf 81,}
  4136--4139 (1998).

\bibitem{norton1}
A.~H. Norton and C.~M. de~Sterke, ``Optimal poling of nonlinear photonic
  crystals for frequency conversion,'' Opt. Lett. {\bf 28,} 188--190 (2003).

\bibitem{hexln}
N.~G.~R. Broderick, G.~W. Ross, H.~L. Offerhaus, D.~J. Richardson, and D.~C.
  Hanna, ``HeXLN: A two dimensional nonlinear photonic crystal,'' Phys. Rev.
  Lett. {\bf 84,} 4345--4348 (2000).

\bibitem{winter}
A. Chowdhury, C. Staus, B.~F. Boland, T.~F. Kuech, and L. McCaughan,
  ``Experimental demonstration of 1535-1555-nm simultaneous optical wavelength
  interchange with a nonlinear photonic crystal,'' Opt. Lett. {\bf 26,}
  1353--1355 (2001).

\bibitem{fib}
S. ning Zhu, Y. yuan Zhu, Y. qiang Qin, H. feng Weng, C. zhen Ge, and N. ben
  Ming, ``Experimental Realization of Second Harmonic Generation in a Fibonacci
  Optical Superlattice of LiTa$O_3$,'' Phys. Rev. Lett. {\bf 78,} 2752--2755
  (1997).

\bibitem{saltiel}
S.~M. Saltiel and Y.~S. Kivshar, ``Phase matching in nonlinear $\chi^{(2)}$
  photonic crystals,'' Opt. Lett. {\bf 25,} 1204--1206 (2000).

\bibitem{fejer1d}
G. Imeshev, M.~A. Arbore, M.~M. Fejer, A. Galvanauskas, M. Fermann, and D.
  Harter, ``Ultra-short pulse second harmonic generation with longitudinally
  nonuniform quasi-phase-matching gratings: pulse compression and shapping,''
  J. Opt. Soc. Am. B {\bf 17,} 304--318 (2000).

\bibitem{fejer2d}
G. Imeshev, M. Proctor, and M.~M. Fejer, ``Lateral patterning of nonlinear
  frequency conversion with transversely varying quasi-phase-matching
  gratings,'' Opt. Lett. {\bf 23,} 673--675 (1998).

\bibitem{fejer2d2}
J.~R. Kurz, A.~M. Schober, D.~S. Hum, A.~J. Saltzman, and M.~M. Fejer,
  ``Nonlinear physical optics with transversely patterned quasi-phase-matching
  gratings,'' IEEE J. Select. Top. Quantum Electron. {\bf 8,} 660--664 (2002).

\bibitem{penrose}
R. Penrose, ``The Role of Aesthetics in Pure and Applied Mathematical
  Research,''  in {\em The Physics of Quasicrystals}, P.~J. Steinhardt and S.
  Ostund, eds., (World Scientific, Singapore, 1987), Chap.~Appendex I.

\bibitem{penrose2}
D. Levine and P.~J. Steinhardt, ``Quasicyrstals. I. Definition and Structure,''
  Phys. Rev. B {\bf 34,} 596--616 (1986).

\bibitem{hexln2}
N.~G.~R. Broderick, R.~T. Bratfalean, T.~M. Monco, D.~J. Richardson, and C.~M.
  de~Sterke, ``Temperature and wavelength tuning of 2nd, 3rd and 4th harmonic
  generation in a two dimensional hexagonally poled nonlinear crystal,'' J.
  Opt. Soc. Am. B {\bf 19,} 2263--2272 (2002).

\end{thebibliography}

\newpage

\newpage

\newpage

\end{document}